\documentclass[proof]{WileyASNA-v1}

\articletype{Original paper}%

\begin{document}

	     \title{New dwarfs around the curly spiral galaxy M\,63}

    \author[1]{I.D. Karachentsev}
    \author[2]{F. Neyer,  R. Sp\"{a}ni, T. Zilch}

    \address[1]{\orgdiv{Russian Academy of Sciences}, \orgname{Special Astrophysical Observatory}, \orgaddress{\state{N.Arkhyz}, \country{KChR, Russia}}}

\address[2]{\orgdiv{Fachgruppe Astrofotografie}, \orgname{Tief Belichtete Galaxien group of Vereinigung der Sternfreunde e.V.}, \orgaddress{\state{PO Box 1169, D-64629, Heppenheim}, \country{Germany}}}

  \corres{*Russian Academy of Sciences, Special Astrophysical Observatory, N.Arkhyz, KChR, Russia \email{ikar@sao.ru}}

\presentaddress{*Russian Academy of Sciences, Special Astrophysical Observatory, N.Arkhyz, KChR, Russia}  
   
\abstract{We present a deep (50 hours exposed) image of the nearby spiral galaxy 
M\,63 (NGC\,5055), taken with a 0.14-m aperture telescope. The galaxy halo 
exhibits the known very faint system of stellar streams extending across 
110~kpc. We found 5 very low-surface-brightness dwarf galaxies around M\,63.
Assuming they are satellites of M\,63, their median parameters are: absolute
$B$-magnitude --8.8~mag, linear diameter 1.3~kpc, surface brightness 
$\sim 27.8$ mag/sq. arcsec and linear projected separation 93~kpc.
Based on four
brighter satellites with measured radial velocities, we derived a low
orbital mass estimate of M\,63 to be (5.1$\pm1.8) 10^{11} M_{\odot}$ on a scale of
$\sim216$~kpc.
The specific property of M\,63 is its declining rotation curve. 
Taking into account the declining rotation curves of the M\,63 and three nearby massive galaxies 
NGC\,2683, NGC\,2903, NGC\,3521, we recognize their low mean 
orbital mass-to-K-band luminosity ratio, $(4.8\pm1.1)~M_{\odot}/L_{\odot}$, that is 
only $\sim1/6$ of the corresponding ratio for the Milky Way and M\,31.}
 
 \jnlcitation{\cname{%
\author{Karachentsev I.D.}, 
\author{Neyer F.}, 
\author{Sp\"{a}ni R.},  and 
\author{Zilch T.}} (\cyear{2020}), 
\ctitle{New dwarfs around the curly spiral galaxy M\,63}, \cjournal{Astronomische Nachrichten}, \cvol{2020}}

 \maketitle

 \section{Introduction.}
  The large spiral Sbc galaxy M\,63 (NGC\,5055), known as the "sunflower" galaxy, has a radial velocity of $V_{LG} = +570\pm3$ km\,s$^{-1}$ relative to the Local Group centroid. Jacobs et al. (2009) and McQuinn et al. (2017) determined its  distance to be 8.99~Mpc and 8.87~Mpc, respectively, using the luminosity of  the tip of the red giant branch method. In the Extragalactic Distance Database  (http://edd.ifa.hawaii.edu) the distance of M\,63 is re-estimated as 9.04$\pm$0.10~Mpc. We will use this value for our estimates below. According to M\"{u}ller et al. (2017), the galaxy M\,63 together with galaxies M\,51 and M\,101 and their companions form a chain of three groups. This filamentary structure may extend further towards the Local Void, including the spiral galaxy NGC\,6503. The radial velocities and distances of principal galaxies in this filament increase smoothly from the northern to the southern edge: $+297\pm13$~km\,s$^{-1}$ (NGC\,6503 at 6.25~Mpc), $+375\pm2$~km\,s$^{-1}$ (M\,101 at 6.95~Mpc), $+553\pm2$~km\,s$^{-1}$ (M\,51 at 8.40~Mpc), and $+570\pm3$~km\,s$^{-1}$ (M\,63 at 9.04~Mpc). 
Hereinafter, the radial velocities and their errors are taken from HyperLEDA ( Makarov et al. 2014, http://leda.univ-lyon1.fr). An angular extension of the filament reaches about 30 degrees.

  At an apparent magnitude of $K_s = 5.60^m$ the M\,63 galaxy has  the extinction corrected luminosity of $L_K = 1.0\times 10^{11} L_{\odot}$, i.e., 2 times higher than that of the Milky Way or M\,31. In the zone of gravitational domination of M\,63 there are 4 dwarf galaxies with similar radial velocities: UGC\,8313 ($V_{LG}$ = +658$\pm$8 km\,s$^{-1}$), KKH\,82 = UGCA\,337 (+545$\pm$26 km\,s$^{-1}$), CGCG\,217-018 
(+608$\pm$33 km\,s$^{-1})$ and SDSS1327+4348 = PGC 2229803 (+529$\pm$39 km\,s$^{-1}$), the luminosities of which are (1/100 -- 1/500)
the luminosity of M\,63. Karachentseva \& Karachentsev (1998) detected three dwarf galaxies of low surface brightness around M\,63: KK\,191, KK\,193, and KK\,194. Their radial velocities remain unmeasured. Using the data on Sloan Digital Sky Survey (SDSS), M\"{u}ller et al. (2017) found three more assumed satellites of M\,63: dw1308+40, dw1305+41, and dw1303+42. Recently Karachentsev 
et al. (2020) detected two new ultra-faint dwarfs near M\,63: TBGdw1 and TBGdw2, inspecting a deep image of M\,63 obtained by O. Schneider with an exposure time of $\sim$10 hours.

  An impressive feature of the M\,63 galaxy is the presence of multiple arches and "plumes" of low surface brightness on its periphery. This faint structure was first noticed by van der Kruit (1979). Staudaher et al. (2015) also found it on their mosaic of deep images of M\,63 exposed in the $K$-band. The system of arches as assumed stellar streams is visible most detailed in the deep B-
and $R$-images of M\,63 presented by Chonis et al. (2011). According to their photometry, a surface brightness of the faintest arches and plumes reaches $SB_B \sim(27-29)$ mag/sq. arcsec and the color of the streams are consistent with the color of stellar population in outer disk and halo of M\,63. It is assumed
that the observed system of streams is most probably a result of accretion of a dwarf satellite in the last few Gyr.

\section{Deep wide-field image of M\,63.}

  The presented image of M\,63 was acquired with the 0.14-m aperture refractor
TEC140ED APO at f/7 focal ratio using a Moravian G3-16200 Monochrom CCD (KAF-16200) camera with Astrodon LRGB filters. 
The mounts used during imaging were an AstroPhysics 900 GoTo (Antares Observatory) and a 10micron GM4000 HPS II (Ceres Observatory).
Guiding was performed using an active off-axis autoguiding system. Exposures with each filter: Lum (18 hours), red (10.7~h), green (9.7~h) blue (12.3~h) with 20-minutes sub-exposures were obtained by F. Neyer in Antares Observatory, Gossau, Switzerland and by R. Sp\"{a}ni (for 9 hours) in Ceres Observatory, Urnasch AR, Switzerland. The total exposure time was 50.7 hours. Image calibration was 
carried out with standard calibration techniques, including: Overscan correction, Dark and Bias correction, Flat-fielding after each imageing session. A master luminance was created by combining all sub-frames of all filters. The total field of view is 120 by 80 arcminutes.

 To address the effect of sky glow, the following procedure was used to calibrate the frames: an individual linear surface gradient (polynomial degree 1) was estimated for each image based on an autonomos distribution of  several thousand background samples. The estimated
gradients were substracted from the images before image integration. Remaining low spatial frequency variations ( > 20 arcmin) across the stacked image were reduced by a few well placed samples, i.e., outside areas with stellar stream features or galactic cirrus. Robust estimates of the individual background samples were used to build a model of low frequency background fluctuations. The model was carefully inspected for systematic artifacts due to galactic cirrus before it was applied.

\begin{figure*}[t]
\centerline{\includegraphics[scale=0.18]{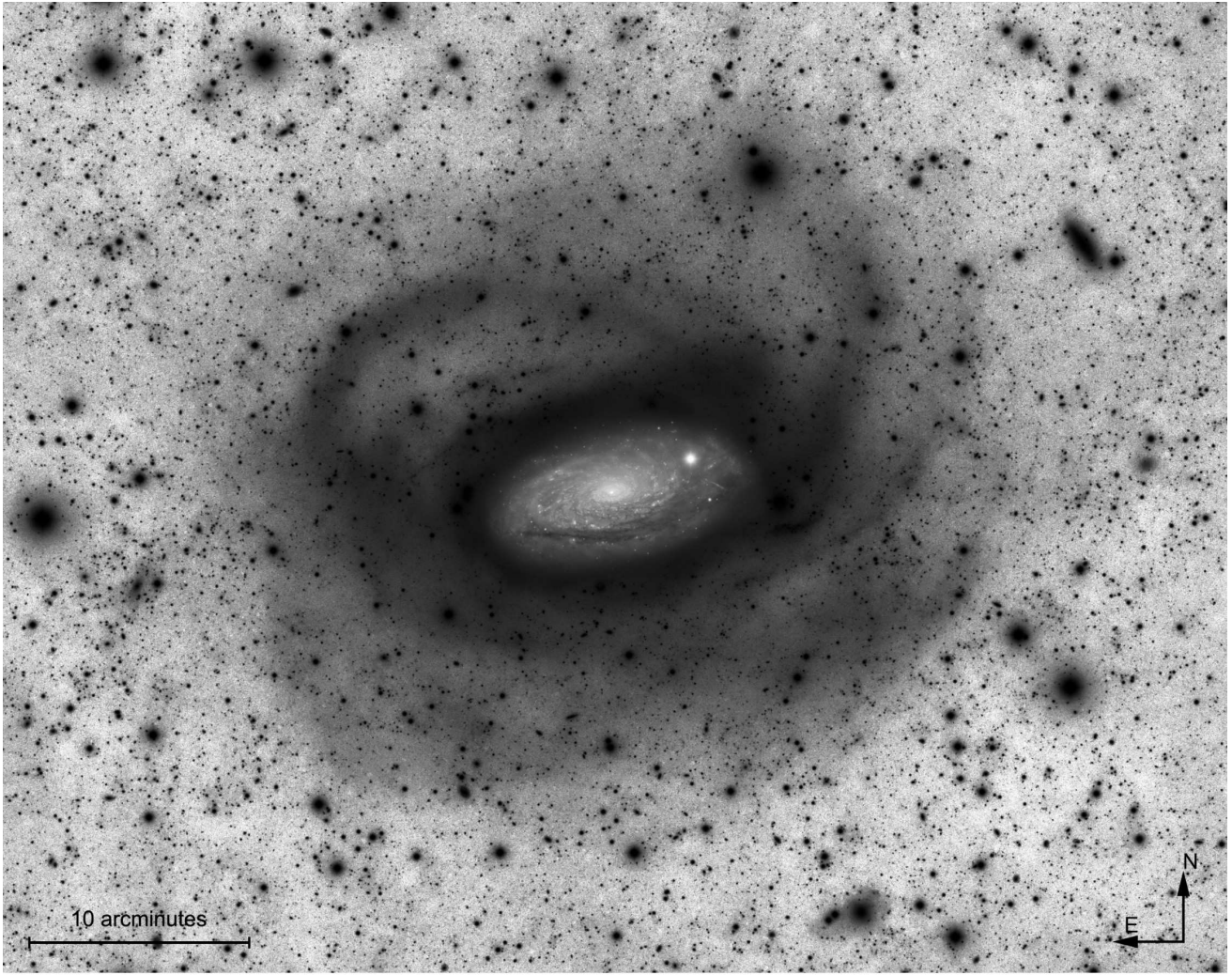}}
\caption{A system of stellar streams around M\,63. The image size is
	  $55\times 44^{\prime}$ or $145\times 116$~kpc.  The scale 10 arcmin is 26.4 kpc. For reference, the 
	  optical image of the galaxy is shown in the central ellipse with
	  the standard major diameter of 32~kpc. North is up and East is on the left. The image is taken by the authors, more details in the text.}
\end{figure*}

  Figure 1 shows the central part of the obtained image (inverted) with the size of 55 by $44^{\prime}$ or 145 by 116~kpc. For reference, the positive image of M\,63's disk has been superimposed to indicate the standard optical diameter of the galaxy  (32 kpc). Figure 1 illustrates a deep view of the faint details of the galaxy periphery showing all the streams shown by Chonis et al. (2011). Our new deep image, however, also demonstrates that the system of arches and loops extends significantly beyond, having a diameter of $\sim110$~kpc. With a rotation amplitude of $V_m \simeq 210$~km\,s$^{-1}$ (Bosma, 1981; Battaglia et al. 2006), the characteristic crossing-time for the system of stellar streams reaches about 1~Gyr. 
Therefore, this image can serve as a suitable archaeological basis for modelling the process of satellite accretion on a time scale that makes up an appreciable fraction of the age of the universe. The direct and inverted images of M\,63 and its surroundings will be available on the website http://tbg.vdsastro.de.

  Inspecting the entire $120\times 80^{\prime}$ size image, we found the known dwarf galaxies:
UGC\,8313, KKH\,82, KK\,191, KK\,193, TBGdw1, and TBGdw2, and also 5 new candidates for M\,63 satellites of very low surface brightness. A mosaic of direct and inverted images of 10 faint and ultra-faint dwarfs is shown in Figure 2.
Each image section has a size of 3.7 by 3.7$^{\prime}$ or 9.7 by 9.7~kpc, assuming the distance of 9.04~Mpc.
\begin{figure*}[t]
\centerline{\includegraphics[scale=0.50]{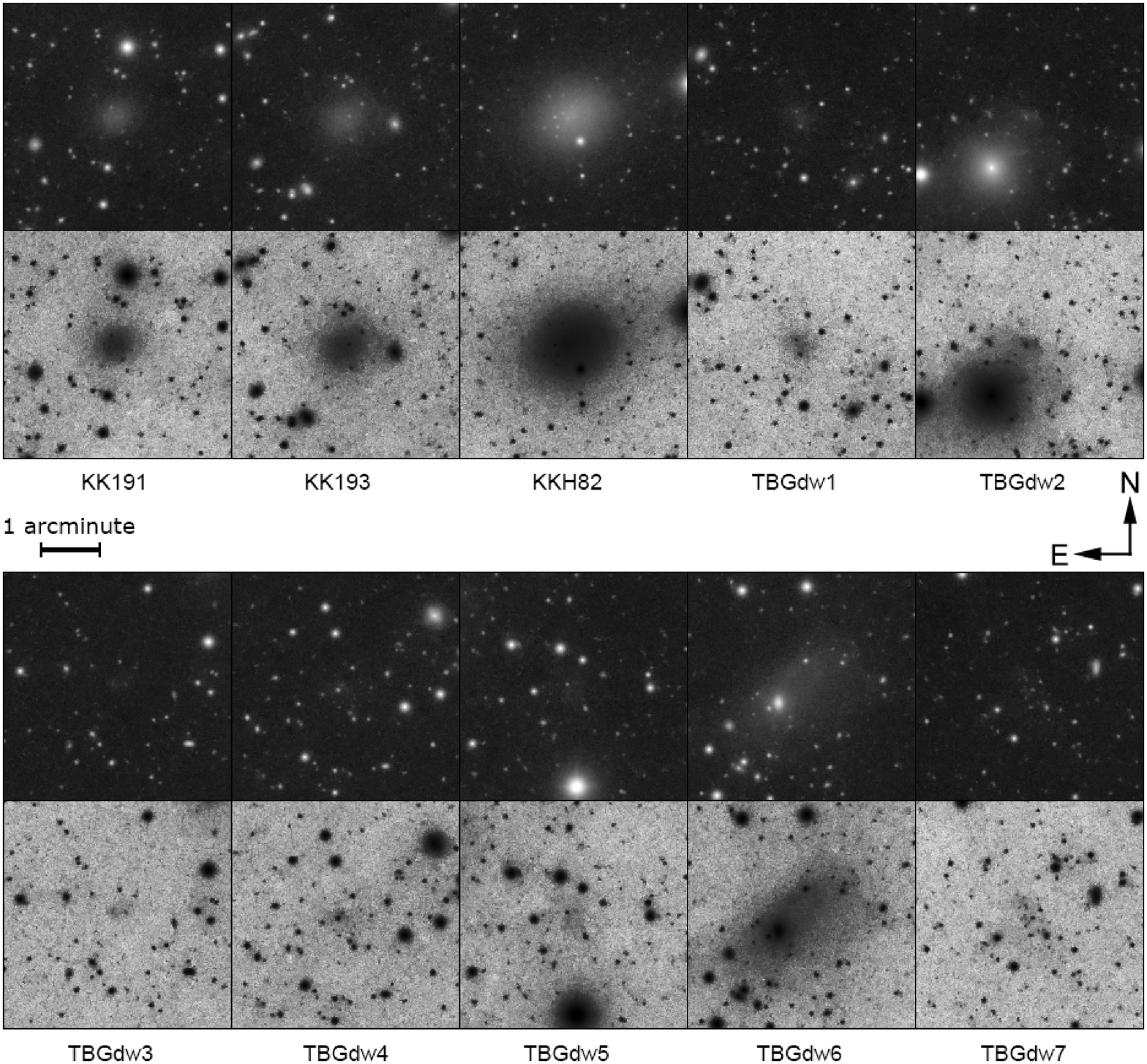}}
\caption{Mosaic of direct and inverted images of low surface brightness dwarf galaxies in the vicinity of M\,63. Each image fragment is spanning $3.7^{\prime}$ across. North is up and East is on the left.}
\end{figure*}

\begin{center}
\begin{table*}[t]%
\centering
\caption{Probable LSB satellites of M\,63 without radial velocities.}
\tabcolsep=0pt%
\begin{tabular*}{17.5cm}{@{\extracolsep\fill}lcclccrcr}
\toprule
 Name     & RA (J2000) DEC  &   SGL   SGB&  Type&$a^{\prime}$&   $B$ & $M_B$ &  $r_p$ &  $R_p$ \\
  (1)     &     (2)         &    (3)&       (4) & (5) & (6)  &  (7)  & (8)  & (9)  \\
\midrule
 TBGdw2 *)& 131447.3+414339 & 76.48 14.02&  Tr  & 0.48&  19.5&  -10.3&  0.36&   57 \\
 KK 191   & 131339.7+420236 & 76.12 13.85&  Tr  & 0.95&  18.2&  -11.6&  0.40&   63 \\
 KK 193   & 131529.8+413011 & 76.72 14.11&  Tr  & 1.07&  18.7&  -11.1&  0.54&   85 \\
 TBGdw5   & 131422.9+413409 & 76.63 13.91&  Sph & 0.42&  20.9&  - 8.9&  0.55&   87 \\
 TBGdw3   & 131258.0+415117 & 76.29 13.70&  Sph & 0.42&  21.6&  - 8.2&  0.56&   88\\
 TBGdw6 *)& 131834.9+414512 & 76.55 14.72&  Irr & 1.80&  18.5&  -11.3&  0.59&   93\\
 TBGdw7   & 131902.8+420023 & 76.31 14.84&  Sph & 0.45&  21.7&  - 8.1&  0.60&   95\\
 TBGdw1   & 131218.5+415833 & 76.15 13.59&  Tr  & 0.52&  19.5&  -10.3&  0.66&  102\\
 TBGdw4   & 131308.5+412736 & 76.70 13.66&  Sph & 0.50&  21.0&  - 8.8&  0.77&  121\\
 dw1308+40& 130846.0+405404 & 77.13 12.76&  Sph & 0.37&  18.4&  -11.4&  1.76&  278\\
 dw1305+41& 130529.0+415324 & 76.02 12.32&  Irr & 0.41&  17.3&  -12.5&  1.94&  306\\
 dw1303+42& 130314.0+422217 & 75.46 12.00&  Irr & 0.38&  18.3&  -11.5&  2.37&  374\\
 KK 194   & 131719.5+442348 & 73.82 14.85&  Irr & 0.63&  17.6&  -12.2&  2.45&  387\\
 dw1305+38& 130558.0+380543 & 79.85 11.73&  Irr & 0.32&  17.9&  -11.9&  4.44&  700\\
\bottomrule
   \end{tabular*}
  \begin{tablenotes}
   \item  *) TBGdw2 locates $\sim1^{\prime}$ NW away from a galaxy WISEAJ131450.79+414247.2
having $m_g = 15.4$~mag and Z = 0.0438. Also, a faint red galaxy
WISEAJ131447.72+414337.4 with Z = 0.6799 is projected into the TBGdw2 body.
 \item *) TBGdw6 may be not a dwarf galaxy but a patch of reflecting nebulae. A
galaxy WISEAJ131836.93+414455.4 with $m_g = 16.9$~mag and Z = 0.06277 is
projected into the TBGdw6.
 \end{tablenotes}
 \end{table*}
 \end{center}
 \begin{center}
 \begin{table*}[h]
  \caption{Assumed members of the M\,63 group with known radial velocities.}
   \tabcolsep=0pt%
  \begin{tabular*}{17.5cm}{@{\extracolsep\fill}lcclrccccr}    
 \toprule
 Name      &    SGL & SGB &  Type&   $B_T$ &  $V_{LG}$ & $r_p$  & $R_p$  \\
\midrule
M 63       & 76.20 &14.25 &  Sbc &  9.3  & {\bf $570\pm3$}     &   0   &  0  \\
UGC 8313   & 75.96 &13.92 & Sm   & 14.7  & {\bf $658\pm8$}     & 0.41  & 65 \\
KKH 82     & 76.36 &13.69 & Tr   & 16.4  & {\bf $545\pm26$}   & 0.58  & 92 \\
CGCG217-018& 77.62 &13.47 & BCD  & 14.9  & {\bf $608\pm33$}   & 1.62  &256 \\
SDSS132753 & 74.65 &16.66 & BCD  & 16.2  & {\bf $529\pm39$}   & 2.87  &453 \\ 
\bottomrule
\end{tabular*}
 \end{table*}

 \end{center}
  Table 1 contains a summary of the known dwarf galaxies of low surface brightness in the vicinity of M\,63, for which radial velocities have not yet been measured. The table columns contain: (1) galaxy name; (2) equatorial coordinates, J2000.0; (3) supergalactic coordinates; (4) morphological type;
(5) apparent $B$-magnitude, estimated for the new objects via their angular diameter and the scale of surface brightnesses from photometry by Chonis et al. (2011); (6) maximum angular diameter in arcmin; (7) absolute $B$-magnitude with a correction of 0.08~mag for the Galactic extinction: (8,9)
angular (in degrees) and linear (in kpc) projected separation of the dwarf from M\,63. The assumed satellites are ranked according to their angular separation from M\,63. The last dwarf galaxy, dw1305+38, found by M\"{u}ller et al. (2017), locates far beyond the virial radius of the M\,63's halo $(R_v \sim 300$~ kpc) and obviously does not match to satellites of M\,63.

\begin{center}
\begin{table*}[t]%
\caption{Nearby luminous galaxies with declined rotation curves.}
\centering
\begin{tabular*}{17.5cm}{@{\extracolsep\fill}llrlllrlr}
 \toprule
 Name      & Type &   D  & $\log(L_K)$&  $n_v$ & $\sigma_{\Delta v}$& $\langle R_p\rangle$ &$\log(M_{\rm orb})$ & $M_{\rm orb}/L_K$ \\
\midrule
 NGC2683   & Sb   & 9.82 & 10.81   & 2    & 43   & 49    & 11.09     &1.9$\pm$1.3 \\
 NGC2903   & Sbc  & 8.87 & 10.82   & 4    & 45   &198    & 11.68     &7.3$\pm$6.4 \\
 NGC3521   & Sbc  &10.70 & 11.09   & 2    & 46   &198    & 11.77     &4.8$\pm$4.0\\
 NGC5055   & Sbc  & 9.04 & 11.00   & 4    & 54   &216    & 11.71     &5.1$\pm$1.8\\
\hline
 Mean      & Sbc  & 9.61 & 10.93   & 3    & 47   &165    & 11.56     &4.8$\pm$1.1\\
\bottomrule
\end{tabular*}
\end{table*}
\end{center}

  Our five new candidates of M\,63 satellites have the median parameters: 
absolute magnitude $M_B = -8.8^m$, linear diameter $\sim1.3$~kpc and the mean 
surface brightness $SB_B\sim27.8$ mag/ sq. arcsec, that are compatible with the
parameters of faint satellites around the Milky Way and M\,31.

\section{Dark halo mass of the M\,63.}
 
  In the vicinity of M\,63, i.e., within a radius of $r_p = 3.0^{\circ} ( = 473$~kpc), four galaxies are known with radial velocities in the range of $\pm300$~km\,s$^{-1}$ relative to the velocity of M\,63. More detailed information is given in Table 2, whereas their morphological types, apparent magnitudes, radial velocities ($V_{LG}$,~km\,s$^{-1}$),
angular ($r_p, ^{\circ}$) and linear ($R_p,$ kpc) projected separations are given. Unfortunately, these galaxies do not have individual distance estimates. On projected separations of 0.5 -- 1.0~Mpc there are 3 more dwarf galaxies with similar radial velocities: CGCG189-050 (664~kpc, +368$\pm5$~km\,s$^{-1}$), WSRT-CVN51
(847~kpc, +573$\pm3$~km\,s$^{-1}$) and DDO\,182 (891~kpc, +730$\pm1$~km\,s$^{-1}$). Individual distance estimates for them are also absent. These remote dwarf galaxies are rather associated with other nearby luminous spirals than with M\,63.

\begin{figure}[t]
\centerline{\includegraphics[scale=0.55]{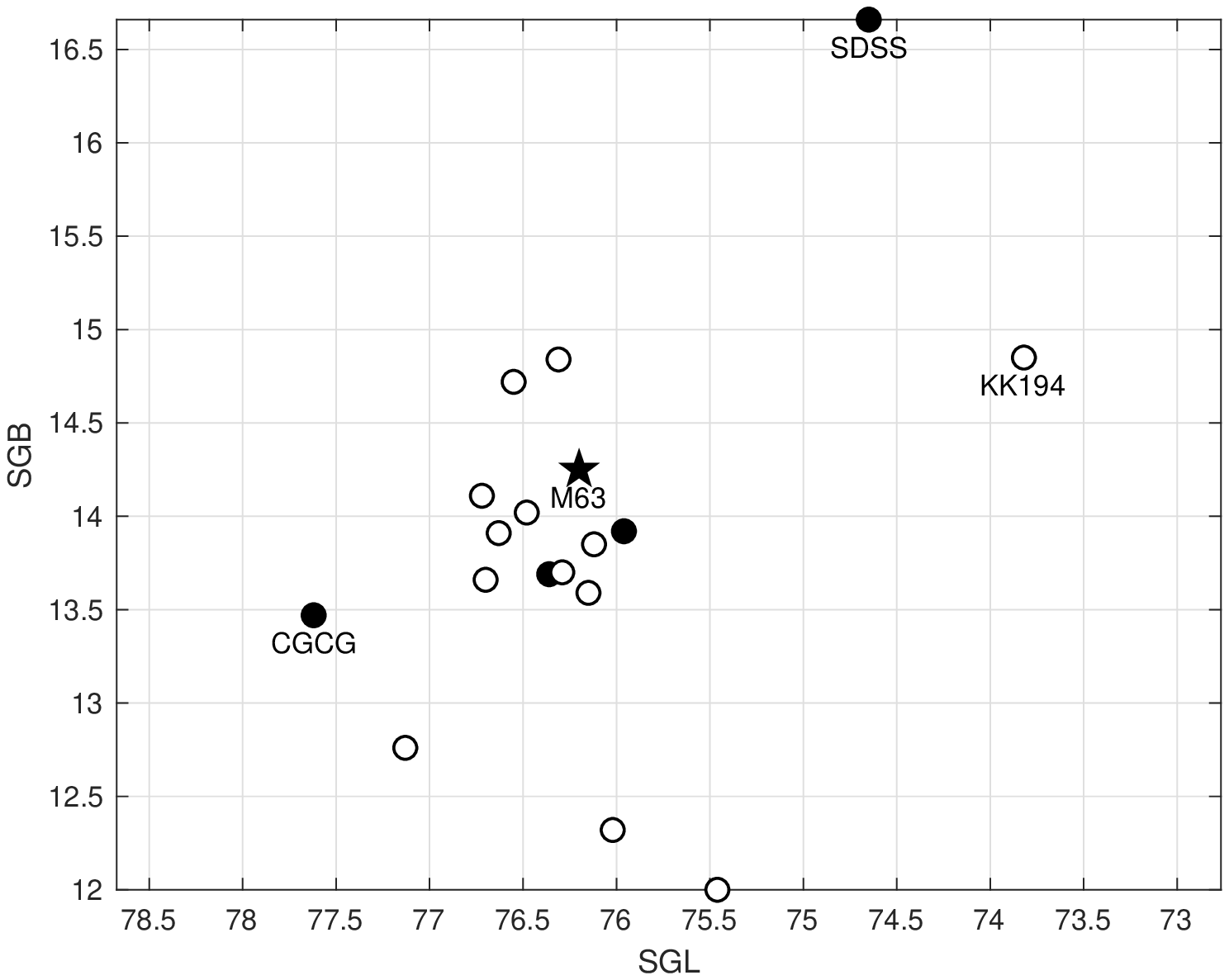}}
\caption{Distribution of dwarf galaxies around M\,63 in supergalactic coordinates. The galaxies with known radial velocities are shown by solid symbols, the low surface brightness dwarfs from Table 1 without radial velocities are indicated by open circles.}
\end{figure}

The distribution of galaxies of the M\,63 group in supergalactic coordinates is presented in Figure 3. Five galaxies with measured radial velocities are indicated with solid symbols and dozen galaxies from Table 1 without radial velocities are shown with open circles. The distribution of low surface brightness satellites looks rather asymmetric relative to M\,63. The reason
for this effect remains unclear.

  The system of 4 satellites of M\,63 is characterized with the average projected separation $\langle R_p\rangle = 216$~kpc and the
radial velocity dispersion $\sigma_{\Delta v} = 54$~km\,s$^{-1}$.  A noticable part of the $\sigma_{\Delta v}$ is caused by velocity measurement errors, $\langle\epsilon_{\Delta v}^2\rangle$ = (26~km\,s$^{-1})^2$.
Assuming an arbitrary orientation of the satellite orbits with the
mean eccentricity $\langle e^2\rangle = 1/2$, the estimate of total (orbital) mass of the central dominated galaxy can be written as

   $$M_{\rm orb} = (16/\pi G) \langle\Delta V^2 R_p\rangle = 1.18\times 10^6 \langle \Delta V^2 R_p\rangle.$$

Here, $G$ is the gravitation constant, and $\Delta V$ and $R_p$ are expressed in km\,s$^{-1}$ and kpc, respectively (Karachentsev \& Kudrya, 2014). For 4 satellites, we obtained an estimate of orbital (virial) mass $M_{\rm orb} = (5.1\pm1.8)\times 10^{11}$~$M_{\odot}$ or the mass-to-luminosity ratio $M_{\rm orb}/L_K = (5.1\pm1.8)$~$ M_{\odot}/L_{\odot}$.  The uncertainty for $M_{\rm orb}$ is estimated via variations
of the product $\Delta V^2 R_p$ from one satellite to another. \footnote{Note, that the average value, $\langle \Delta V^2 R_p\rangle $, gives a
statistically biased mass estimate. In the presence of large radial velocity errors, $\epsilon_ {\Delta v}$, the product 
 $\langle \Delta V^2 R_p\rangle $ should be replaced by $\langle(\Delta V^2 - \epsilon^2_{\Delta V}) R_p\rangle $, decreasing the orbital mass estimate.} 
The derived ratio of $M_{\rm orb}$ is significantly lower than that of ($27\pm9)$~$M_{\odot}/L_{\odot}$ for the Milky Way and $(33\pm6)$~$M_{\odot}/L_{\odot}$ for M\,31 (Karachentsev \& Kudrya, 2014). We are not aware of other examples of large galaxies from the literature with such a low dark-to-stellar mass ratio.

  Battaglia et al. (2006) investigated the kinematics of M\,63 using observations in the 21-cm line with the Westerbork Synthesis Radio Telescope. Based on the results of these observations they built the rotation curve of the galaxy, V(R), which has a steep rise with the maximum at $\sim$10 kpc and than a smooth velocity decreasing down to $\sim175$~km\,s$^{-1}$ within the last measured point at the distance of 50~kpc (at D = 9.04~Mpc). To fit the observed $V(R)$, the authors constructed the galaxy model with contribution of different mass components: a warped gaseus disk, stellar disk and dark matter halo. The best agreement of the model with observed
circular velocities was achieved at $M_{\rm gas} = (0.1\times 10^{11})$~$M_{\odot}, M_{\rm star} = 
(1.0\times 10^{11})$~$M_{\odot}$ and $M_{DM} = (2.4\times 10^{11})$~$M_{\odot}$. Thus, the total mass estimate within 50~kpc amounts to $(3.5\times 10^{11})$~$M_{\odot}$ that is more than half the total mass we estimated when using the satellite motion at a scale of $\sim$216~kpc. Consequently, the galaxy M\,63 has a rather skinny dark matter halo.

  Casertano \& van Gorkom (1991) noted that among the Local Volume galaxies with distances $D < 11$~Mpc there are two galaxies, NGC\,2683 and NGC\,3521, with decreasing rotation curves at their periphery. Recently, Zobnina \& Zasov (2020) presented a list of 22 galaxies with decseasing $V(R)$. Besides the luminous galaxies of the Local Volume, M\,63 and NGC\,3521, this list contains one more isolated nearby galaxy of high luminosity: NGC\,2903. Using the data of Updated Nearby Galaxy Catalog (Karachentsev et al. 2013), we looked for dwarf satellites of these galaxies within $R_p = 500$~kpc in the velocity range of $\pm300$~km\,s$^{-1}$. The results are presented in Table 3. As there are no bright satellites of the Magellanic Clouds type around these galaxies, their number is small. However, it can be stated that the available faint satellites have a low velocity dispersion, $\sigma_{\Delta v}= ~47$~km\,s$^{-1}$. This leads to the average orbital mass-to-luminosity ratio of $\langle M_{\rm orb}/L_K\rangle = (4.8\pm1.1)$~$M_{\odot}/L_{\odot}$ on the scale of $\langle R_p\rangle = 165$~kpc.

  It is curious to note that the galaxy NGC\,3521, like M\,63, has a strongly disturbed periphery with an arcuate structure of low surface brightness (Karachentsev et al. 2020).

\section{Concluding remarks}

  A deep image of M\,63, obtained with an amateur 0.14-m aperture telescope and an exposure time of 50 hours, shows that the system of faint stellar streams at the periphery of this galaxy extends up to 55~kpc from its center. Obviously, these streams are the remnants of the accretion process of one or several satellites of moderate mass. In addition to the already known satellites of M\,63, we found 5 more candidates for fainter satellites with a median absolute magnitude of $-8.8$~mag, a median linear diameter of $\sim1.3$~kpc and the mean surface brightness $\sim27.8$~mag/sq. arcsec.

  From four satellites of M\,63 with measured radial velocities, we estimated the orbital mass of the M\,63 halo and obtained the ratio of the total mass-to-stellar luminosity of $M_{\rm orb}/L_K = (5.1\pm1.8)$~$M_{\odot}/L_{\odot}$. This value is 5 -- 6 times
lower than the respective ratio for the Milky Way and M\,31.

 Unlike most spiral galaxies, the M\,63 galaxy is distinguished by the presence of decreasing (Keplerian) region on its rotation curve. Several other nearby luminous spiral galaxies (NGC\,2683, NGC\,2903, and NGC\,3521) have the same property. They all show a tendency to have poor, low-mass dark matter haloes with the
mean ratio of $\langle M_{\rm orb}/L_K\rangle = (4.8\pm1.1)$~$M_{\odot}/L_{\odot}$. These galaxies of big stellar mass but with a small number of faint satellites, may form a special
population of spiral galaxies characterized by underdeveloped dark halos.
The search for new faint satellites around this type galaxy and the measurement of their radial velocities is an urgent observational problem.

\section*{Acknowledgments}
 We thank the referee for thoroughly reading the manuscript and valuable comments. IDK was supported by RFBR grant 18--02--00005.

\bigskip
{\bf References}
 
Battaglia G., Fraternali F., Oosterloo T., Sancisi R., 2006, A \& A, 447, 49

Bosma A., 1981, AJ, 86, 1791

Casertano S., van Gorkom J.H., 1991, AJ, 101, 1231

Chonis T.S., Martinez-Delgado D., Gabany R.J., et al, 2011, AJ, 142, 166

Jacobs B.A., Rizzi L., Tully R.B., et al, 2009, AJ, 138, 332

Karachentseva V.E., Karachentsev I.D., 1998, A \&AS, 127, 409

Karachentsev I.D., Riepe P., Zilch T., 2020, Ap, 63, 5

Karachentsev I.D., Makarov D.I., Kaisina E.I., 2013, AJ, 145, 101

Karachentsev I.D., Kudrya Y.N., 2014, AJ, 148, 50

 Makarov D.I., Prugniel P., Terekhova N., et al, 2014, A \& A, 570A, 13

McQuinn K.B.W., Skillman E.D., Dolphin A.E., et al, 2017, AJ, 154, 51

M\"{u}ller O., Scalera R., Binggeli B., Jerjen H., 2017, A \& A, 602A, 119

Staudaher S.M., Dale D.A., van Zee L., et al, 2015, MNRAS, 454, 3613

van der Kruit, P.C. 1979, A\&AS, 38, 15
 
 Zobnina D.I., Zasov A.V., 2020, Astronomy Reports, 64, 295


 \end{document}